\documentclass[reprint,prc,
 amsmath,amssymb,
 aps,
 lengthcheck,
floatfix,
]{revtex4}
\usepackage{tabularx}
\usepackage{booktabs}
\usepackage{balance}
\usepackage{graphicx}
\usepackage{dcolumn}
\usepackage{bm}
\usepackage{hyperref}
\usepackage[utf8]{inputenc}
\usepackage{amsmath}
\usepackage{multirow}
\usepackage{soul}
\usepackage{float}
\usepackage{graphicx}
\usepackage{times}
\usepackage[normalem]{ulem}
\usepackage{color}
\usepackage{cellspace}
\usepackage{color}

\usepackage{lipsum}

\newcommand{\be}{\begin{equation}}
\newcommand{\ee}{\end{equation}}
\newcommand{\bea}{\begin{eqnarray}}
\newcommand{\eea}{\end{eqnarray}}

\newcommand {\nonu}{\nonumber}

\newcommand{\comment}[1]{}
\renewcommand\sout{\bgroup \color{red} \ULdepth=-.5ex \ULset}
\def\simge{\mathrel{\rlap{\raise 0.511ex
     \hbox{$>$}}{\lower 0.511ex \hbox{$\sim$}}}}
\def\simle{\mathrel{\rlap{\raise 0.511ex
      \hbox{$<$}}{\lower 0.511ex \hbox{$\sim$}}}}

\begin{document}
\title{Calibrating global behaviour of equation of state by combining nuclear \\ and astrophysics inputs in a machine learning approach}
\author{\href{https://orcid.org/0000-0003-3308-2615}Sk Md Adil Imam$^{1,2}$\includegraphics[scale=0.06]{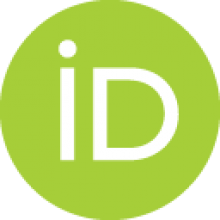}}
\email{mdadil.imam@saha.ac.in}
\author{\href{https://orcid.org/0000-0002-4997-6544}Prafulla Saxena$^{3}$\includegraphics[scale=0.06]{Orcid-ID.png}}
\email{prafulla1308@gmail.com}
\author{\href{https://orcid.org/0000-0003-2633-5821}Tuhin Malik$^4$\includegraphics[scale=0.06]{Orcid-ID.png}
}

\author{\href{https://orcid.org/0000-0003-0103-5590} N. K. Patra
$^5$\includegraphics[scale=0.06]{Orcid-ID.png}
}
\author{\href{https://orcid.org/0000-0001-5032-9435}B. K. Agrawal$^{1,2}$\includegraphics[scale=0.06]{Orcid-ID.png}
}
\email{bijay.agrawal@saha.ac.in}
\affiliation{$^1$Saha Institute of Nuclear Physics, 1/AF 
Bidhannagar, Kolkata 700064, India}  
\affiliation{$^2$Homi Bhabha National Institute, Anushakti Nagar, Mumbai 400094, India}
\affiliation{$^3$Malaviya National Institute of Technology, Jaipur, India}
\affiliation{$^4$CFisUC, Department of Physics, University of Coimbra,
3004-516 Coimbra, Portugal}

\affiliation{$^5$Department of Physics, BITS-Pilani, K. K. Birla Goa Campus, Goa 403726, India}

\date{January 2024}

\begin{abstract}
We implemented symbolic regression techniques to identify suitable analytical functions that map various
properties of neutron stars (NSs), obtained by solving the Tolman-Oppenheimer-Volkoff (TOV) equations, to a few key parameters of the equation of state (EoS). These symbolic regression models (SRMs) are then employed to perform Bayesian inference with a comprehensive dataset from nuclear physics experiments and astrophysical observations. The posterior distributions of EoS parameters obtained from Bayesian inference using SRMs closely match those obtained directly from the solutions of TOV equations. Our SRM-based approach is approximately 100 times faster, enabling efficient Bayesian analyses across different combinations of data to explore their sensitivity to various EoS parameters within a reasonably short time.
\end{abstract}

\maketitle

\section{Introduction} 
Understanding the behavior of matter at densities beyond those found in atomic nuclei is crucial for advancements in modern astrophysics and nuclear physics. The dead remnants of stellar evolution, neutron stars (NS) are perfect laboratories for studying particles in dense environments as their central densities go up to 5-8 times the nuclear saturation density ($\rho_0\sim$ 0.16 fm$^{-3}$). So multi-messenger observations related to NS properties such as gravitational mass, radius, and tidal deformability, provide a way to study the matter at extreme conditions. Observations from massive radio pulsars \cite{Fonseca2021,Romani2022}, the multimessenger event GW170817 \cite{Abbott2017,Abbott2019}, observations of Neutron Star Interior Composition Explorer (NICER)\cite{Riley2019,Miller2019,Riley:2021pdl,Miller:2021qha} of two pulsars, terrestrial endeavors such as heavy ion collision experiments \cite{Akiba2015,Meehan2016,HADES2022,Senger2021}, and theoretical calculations based on chiral effective field theory \cite{Hebeler2015,Lynn2019,Drischler2021a,Drischler2021b} have played crucial roles in deciphering the global behavior of the Equation of State (EoS). 

Future, observations of coalescing BNS events by detectors like LIGO-Virgo-KAGRA, Einstein Telescope~\cite{Punturo2010}, and Cosmic Explorer~\cite{Reitze2019} are likely to occur more frequently, enabling a more precise determination of the high-density part of the EoS~\cite{Most2018,Traversi2020,Raaijmakers2021,Annala2022,Ghosh2022,Iacovelli2023,Huxford2023,Rose2023,Pradhan2023,Pradhan2023b,Walker2024,Huang2024,Patra2023b}. The EoS at low densities $\rho<2\rho_0$ are constrained by various nuclear physics experiments measuring the bulk properties of finite nuclei such as the nuclear masses, neutron skin thickness in $^{208}$Pb, the dipole polarizability and isobaric analog states as well as the heavy-ion collisions. Usually one utilizes these data from nuclear physics experiments and astrophysical observations to constrain the EoS through a Bayesian approach\cite{Tews2012,Hebeler2013,LeFevre2015,Russotto2016,Lynn2016,Monitor2017,Coughlin2017,Drischler2017,Zhang2018,Drischler2020,Dietrich2020,Raaijmakers2020,Huth2021,Biswas2021,Tsang2024,Imam2024b,Malik2024b}. Moreover, one needs to perform the analysis using different combinations of these dataset to asses their importance in determining various key parameters of a EoS. The Bayesian approach utilizing the likelihood functions yielding the posterior distributions for various NS observable is generally time-consuming~\cite {Bilby_ref,Riley2018,Brandes2023,Tiwari2024}. This is primarily due to the computation of NS properties through the solutions of the Tolman-Oppenheimer-Volkoff (TOV) equations. To overcome this issue recently there have been few attempts to map the EoS to the NS properties so that the NS properties can be calculated without recourse to the solutions of TOV equations~\cite{Soma2023, Richter2023,Imam2024,Divaris2024,Carvalho2024,ofengeim2024,Reed2024b,Kumar2024}. 

In Ref. \cite{Imam2024}, we have performed a systematic analysis and correlation studies involving multiple parameters to identify the essential nuclear matter parameters crucially influencing the tidal deformability of neutron stars derived from solutions of TOV equations for neutron star masses ranging from 1.2 - 1.8 M$_\odot$. Here, we demonstrate the mapping of neutron star properties to the key nuclear matter parameters. Our established empirical relation allows for a direct application in facilitating Bayesian analysis for diverse astrophysical observations. In Ref. \cite{Richter2023}, three cutting-edge methodologies for constructing regression models-bidirectional stepwise feature selection, LASSO regression, and neural network regression are utilized. They are applied to pinpoint the crucial EoS parameters influencing the radius R$_{1.4}$ of canonical NSs. This analysis is based on the posterior EoSs derived from Bayesian analyses of observational data on NS. All three regression techniques indicate that only a few parameters governing the NS EoS are significant predictors of R$_{1.4}$.

In this study, we refrain from limiting our parameter space, in contrast to the approach outlined in Ref.\cite{Imam2024}. Instead, we employ a symbolic regression technique to construct analytical expressions of neutron star properties in terms of a few key EoS parameters. In Ref. \cite{Imam2024}, we focused solely on constructing functions for the tidal deformability of NS at some specific masses. However, in the present investigation, we have developed analytical expressions for NS maximum mass,  radii, and tidal deformabilities across a broad spectrum of masses ranging from 1.1 to 2.15 M$_\odot$. These expressions are then incorporated into a Bayesian framework, utilizing data from nuclear physics experiments and astrophysical observations, to probe the overall behavior of the EoS.

This paper is organized as follows. In Sec~\ref{meth} the construction of EoSs, symbolic regression method and Bayesian framework are briefly described. The results are discussed in Sec.\ref{res}. The conclusions are drawn in Sec.\ref{conc}. Finally, the equations obtained using symbolic regression are provided in the Appendix for future use.

\section{Methodology}\label{meth}
Symbolic regression, a machine learning approach is utilized to map the NS properties to a few key EoS parameters. The EoS are constructed using $\frac{n}{3}$ expansion as detailed in \cite{Lattimer2016,Gil2017,Imam2022}. 
\subsection{EoS for NS Matter}\label{cons_EoS}
We constructed the EoS for the core region of NS by employing a $\frac{n}{3}$ expansion within the density range of 0.5$\rho_0$ to 8$\rho_0$. For densities below 0.5$\rho_0$, the neutron star is characterized by an outer and inner crust. The EoS for the outer crust, covering up to 0.00016$\rho_0$, is derived from \cite{Baym1971}, while a polytropic form of the EoS, described in \cite{Carriere2003}, is chosen for the inner crust. These EoSs are utilized in computing neutron star properties through the solutions of the Tolman-Oppenheimer-Volkoff equations \cite{Oppenheimer:1939ne,Tolman:1939jz,Hinderer_2008}. The chosen EoS satisfy conditions of : (i)thermodynamic stability, (ii)positive semi-definiteness of symmetry energy, (iii)causality of the speed of sound, and (iv) maximum mass of stable non-rotating neutron stars, M$_{max}\geq$ 2.15 M$_\odot$.

\subsection{Symbolic Regression}\label{symbolic}
Symbolic regression is a type of regression analysis that aims to discover the best mathematical expression or formula out of all possible equations representing the relationship between input and output variables~\cite{Koza1992,Schmidt2009,Sheng2019,Makke2023}. This process is beneficial when the underlying relationship between variables is unknown or complex. The process involves searching through a vast space of mathematical expressions, typically represented as combinations of mathematical operators (e.g., addition, subtraction, multiplication, division) and a wide range of functions (e.g., sin, cos, exp, log), and can also include user-defined functions, constants, or conditional statements. This search is often performed using various optimization algorithms. While genetic programming and evolutionary algorithms are popular choices due to their ability to efficiently explore large search spaces, other techniques like simulated annealing or gradient-based methods can also be employed, depending on the specific problem and computational constraints. To perform symbolic regression, one has to provide the features (input variables) and their corresponding target values (output variables). A K-fold cross-validation procedure is often used to assess model performance and avoid overfitting. This procedure yields multiple equations in each of the 'K' folds. To choose the best equation, the goodness of fit and model complexity are good metrics to filter out among various searched equations. Simpler models are often preferred to avoid overfitting, even with slightly lower goodness-of-fit scores. The goodness of fit evaluates how well a statistical model, such as a regression equation, fits the observed data. Several metrics are commonly used for this purpose, two of them are (i) R$^2$ which is defined as
\bea
R^2 &=& 1-\frac{SS_{residual}}{SS_{total}}\label{Rsq}\\
 SS_{residual} &= & \sum (y_i - \hat{y}_i)^2 \\
 SS_{total} &= &\sum (y_i - \bar{y})^2. \\ \nonu
 \eea
 Here y$_i$ is the observed or actual value, $\hat{y}_i$ is the predicted value for the i$^{th}$ data point and $\bar{y}$ is the mean of the 'n' observed values. R$^2$ measures the proportion of the variance in the dependent variable that is explained by the independent variables in the model. It ranges from 0 to 1, with higher values indicating a better fit.\\
(ii) RMSE which is defined as,
\bea
{\rm RMSE} = \sqrt{\frac{SS_{residual}}{n}}\label{RMSE}
\eea
This also provide information about the accuracy of a regression model, with lower values indicating a better fit.

\subsection{Bayesian Inference}
The Bayesian likelihood is a fundamental concept in Bayesian statistics, providing a way to update the probability estimate for a hypothesis as more evidence or information becomes available. In the context of Bayesian inference, the posterior distribution measures the plausibility of a set of parameter values given the observed data.

{\it Bayes' Theorem:-} Bayes' theorem is used to update the probability estimate of a hypothesis as more evidence is available. It is formulated as:
\begin{equation}
    P(\theta | {\mathcal D}) = \frac{{\mathcal L}(\mathcal D | {\theta}) P(\theta)}{P({\mathcal D})}
\end{equation}

In the above  equation:
\begin{itemize}
    \item \(P(\theta | {\mathcal D})\) is the posterior probability of the parameters $\theta$ given the data, \(\mathcal D\).
    \item ${\mathcal L}(\mathcal D | {\theta})$ is the likelihood of the data under the parameters (likelihood function).
    \item \(P(\theta)\) is the prior probability of the parameters $\theta$.
    \item \(P({\mathcal D})\) is the marginal likelihood or evidence, often acting as a normalizing constant.
\end{itemize}
For the details about the likelihood function see the Bayesian inference section of Ref.\cite{Agathos2015,Wysocki2020,Landry2020PhRvD,Biswas2020puz,Imam2024b}.

\section{Results and Discussions}\label{res}
{\renewcommand{\arraystretch}{1.35}
\begin{table}[H]
\caption{Prior distributions of the NMPs (\(\theta\)). Here $\rho_0$ = 0.16 fm$^{-3}$, e$_0$ = -16.0 MeV. All other parameters are uniformly distributed between a minimum (\(\theta_{min}\)) and a maximum (\(\theta_{max}\)). For the symbolic regression all the NMPs are reduced to the standard variable, $\hat{\theta} = \frac{(\theta -\bar{\theta})}{\sigma_\theta}$ with mean \(\bar{\theta}\) and standard deviation \(\sigma_\theta\). The unit of all the NMPs are in MeV. } \label{tab1}
  \centering
 
  \begin{tabular}{ccccccccc}
  \hline
 & {K$_0$} & Q$_0$ & Z$_0$ & {J$_0$} &{L$_0$} & {K$_{\rm sym,0}$} & {Q$_{\rm sym,0}$} & {Z$_{\rm sym,0}$} \\ [1.3ex]
 \hline
 \(\theta_{min}\) & 180 & -1500 & -3000 & 25  & 20 & -300& 300 & 3000\\[1.3ex] 
 \hline
 \(\theta_{max}\) & 330 & 1500 & 3000 & 40 & 120 & 300 & 1500 & -3000\\
 [1.3ex] 

 \hline
\(\bar{\theta}\)       & 256.54 & -0.52 & -1049.42 & 31.99 & 76.04 & -41.12 & 987.38 & 200.38 \\[1.3ex] 
 \hline
 \(\sigma_\theta\) & 48.99  & 266.00   & 1558.57  & 4.28  &  27.57 & 154.15 & 331.14 & 1694.79 \\[1.3ex] 
 \hline
  \end{tabular}
\end{table}

\begin{figure*}
    \centering
    \includegraphics[width=0.85\textwidth]{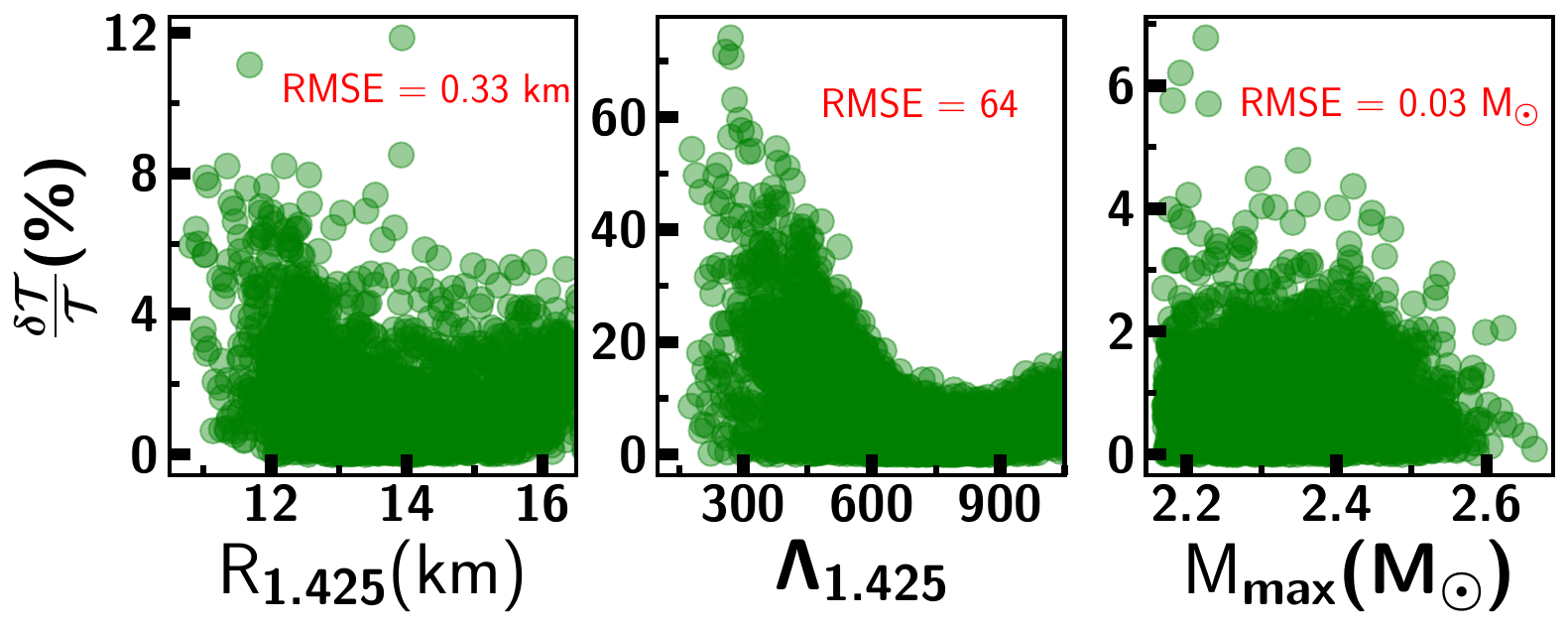}
    \caption{Prediction of symbolic regression models for radius, tidal deformability of NS of mass 1.425 M$_\odot$ and maximum mass of NS. The root mean squared error (RMSE) is also quoted in each panel.}
    \label{fig1}
\end{figure*}
We construct the Symbolic Regression Models(SRMs) to express the NS maximum mass, radii and tidal deformabilities in terms of various nuclear matter parameters as briefly outlined in Sec~\ref{symbolic}. Once SRMs are obtained using the solutions of the TOV equations, they can facilitate constraining the EoS through Bayesian inference. We first validate SRMs for various NS properties considered. Next, we study the sensitivity of the EoS to various nuclear physics and astrophysical inputs considered. Finally, we compared the results from Bayesian inference for a realistic set of data with SRMs and with those obtained directly from the solutions of TOV equations. 
\subsection{Construction and Validation of the SRMs}\label{BT_SRMs}
A large number ($\sim$ 10$^4$) of EoSs are generated for the neutron star matter as described in Sec~\ref{cons_EoS}. The NMPs required for the EoSs are drawn randomly from their uniform distributions (see Table~\ref{tab1}). The average (\(\bar{\theta}\)) and standard deviation (\(\sigma_\theta\)) of each NMPs(\(\theta\)) listed in Table~\ref{tab1} are obtained for the valid EoSs which satisfy the physical conditions and lower bound on NS maximum mass as stated in Sec~\ref{cons_EoS}.
The NS properties obtained for these EoSs through the solutions of the TOV equations are then fitted to the NMPs through symbolic regression to construct the SRMs. For the purpose of symbolic regression first all the features (NMPs), $\theta$ are reduced to standardized form $\hat{\theta} = \frac{(\theta -\bar{\theta})}{\sigma_\theta}$, so that each of them can be treated equally. We have modelled NS radii and tidal deformabilities with masses between 1.1 - 2.15 M$_\odot$ and maximum mass (M$_{\rm max}$) of NS in terms of the nuclear matter parameters directly through the symbolic regression. We employed symbolic regression to explore a diverse set of potential equations, aiming to identify the most optimal mathematical representation. We allowed basic elementary mathematical operations along with more complex operations like square, cube, square root, inverse, sine, cosine, and tangent. This approach allows for a comprehensive search across various functional forms, ensuring the best possible model fit. The expression of the NS property,${\mathcal T}$ takes the form :
 \bea
 {\mathcal T}_{\rm SRM} &=& a_0 + \sum_i a_i \hat{\theta_i} +  \sum_i \sum_j a_{ij} \hat{\theta_i}\hat{\theta_j}\label{eq1}
 \eea
 where, $\theta_i \in$ \{K$_0$, Q$_0$, Z$_0$, J$_0$, L$_0$, K$_{\rm sym0}$, Q$_{\rm sym0}$, Z$_{\rm sym0}$\}
 with a$_0$, a\(_i\) and a\(_{ij}\) are the coefficients of the equations obtained through the cross-validation method in symbolic regression. The equations are selected based on a balance between achieving a high R$^2$ (Eq~\ref{Rsq}), low RMSE (Eq~\ref{RMSE}), and less model complexity. These equations are provided in the Appendix. Using these equations we compute the radius and tidal deformability of neutron stars with masses ranging between 1.1 - 2.15 M$_\odot$ in the interval of 0.05 M$_\odot$ for which we have the analytical expressions from symbolic
 \begin{figure}
    \centering
    \includegraphics[width=0.5\textwidth]{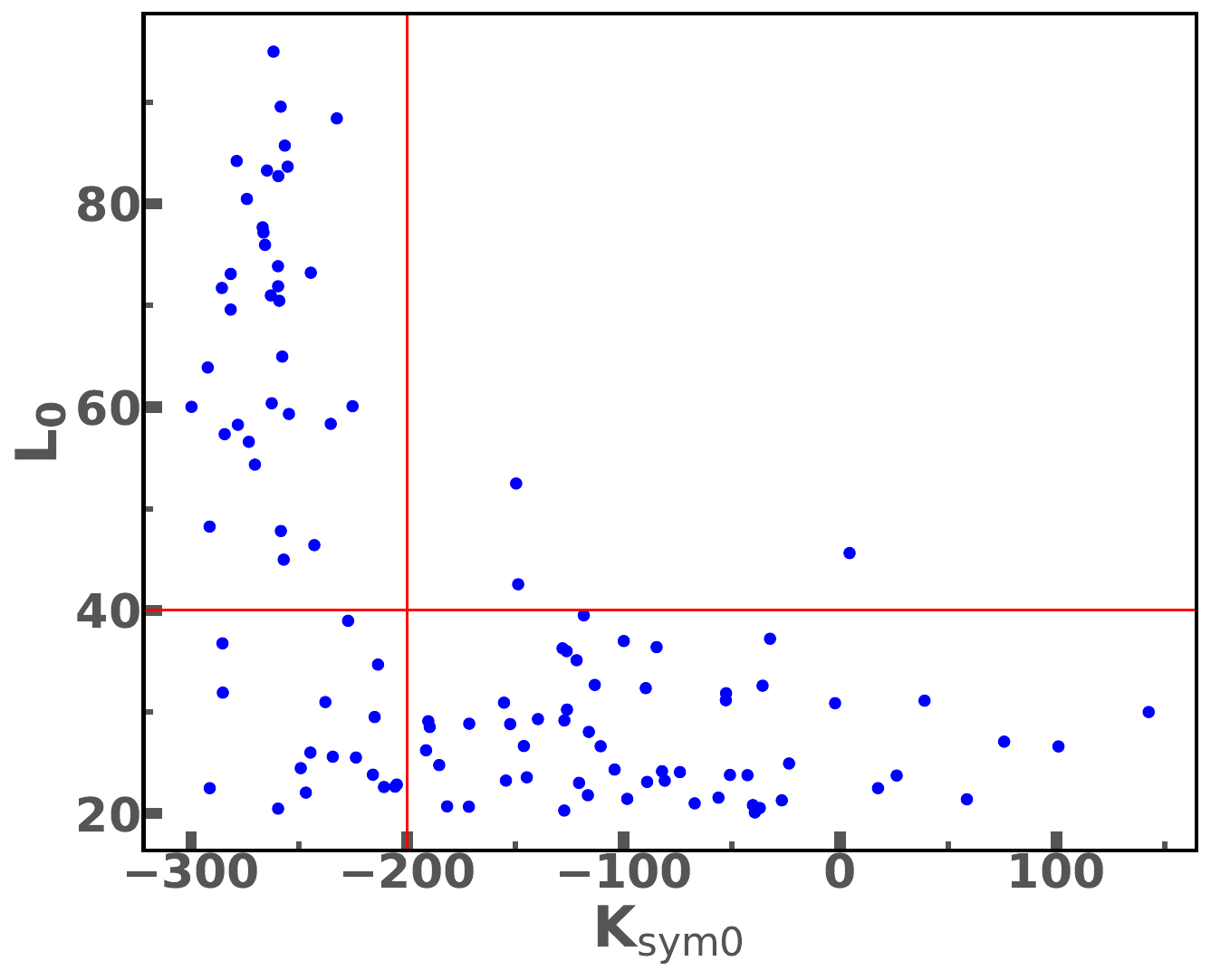}
    \caption{The distribution of L$_0$ (in MeV) and K$_{\rm sym0}$ (in MeV) for the cases for which symbolic regression model predictions yield, $\frac{\delta\Lambda}{\Lambda}>$0.3(see Fig~\ref{fig1}) for tidal deformability of 1.425 M$_\odot$ NSs.}
    \label{fig2}
\end{figure}
 {\renewcommand{\arraystretch}{1.5}
\begin{table*}

  \caption{Pseudo data on a few of the NS properties and key EoS quantities generated for a given set of NMPs listed as `ref' in Table~\ref{tab3}.
  The NS properties considered are radii (R$_{\rm M}$ in km), dimensionless tidal deformabilities($\Lambda_{\rm M}$),  maximum mass (M$_{max}$ in M$_\odot$) with uncertainties 6$\%$, 30$\%$ and 2$\%$ respectively. The key EoS quantities are symmetry energy (e$_{\rm sym}$($\rho$)), symmetry energy pressure (P$_{\rm sym}$($\rho$)), pressure due to snm (P$_{\rm snm}$($\rho$)) referred as ``pseudo experimental data" and energy of the pure neutron matter (e$_{\rm pnm}$($\rho$)), slope of incompressibility (${\mathcal M}$($\rho$)) at the density $\rho$ fm$^{-3}$. The uncertainties on e$_{\rm pnm}$ and ${\mathcal M}$ are 0.5 MeV and 100 MeV, respectively. The uncertainties on e$_{\rm sym}$($\rho$), P$_{\rm sym}$($\rho$), P$_{\rm snm}$($\rho$) are same as in Table-I of Refs~\cite{Imam2024b}. The units of all the quantities related to the EoS are MeV.}
  \label{tab2}
  
  \begin{tabularx}{\textwidth}{|>{\centering\arraybackslash}m{3cm}|>{\centering\arraybackslash}m{3cm}|>{\centering\arraybackslash}m{3cm}|>{\centering\arraybackslash}m{3cm}|}
  
    \cline{1-4}
    \multicolumn{2}{|c|}{\textbf{NS properties}} & \multicolumn{2}{c|}{\textbf{EoS quantities}} \\
    \cline{1-4}
    R$_{1.4}$ & 13.24 & e$_{pnm}$(0.1) & 10.43 \\
    & & ${\mathcal M}$(0.1) & 906 \\
    & & e$_{sym}$(0.035) & 12.86 \\
    R$_{2.08}$ & 12.65  & e$_{sym}$(0.05) & 15.77\\
     &  & e$_{sym}$(0.069) & 19.20 \\
     & &  e$_{sym}$(0.101) & 24.5 \\  
    $\Lambda_{1.4}$ & 597 & e$_{sym}$(0.106) & 25.28 \\  
    & &  e$_{sym}$(0.115) & 26.66 \\

    & &  e$_{sym}$(0.232) & 42.88 \\
    
    M$_{max}$ & 2.27 & P$_{sym}$(0.11) & 1.86 \\
    
     &   & P$_{sym}$(0.232) & 7.25 \\
     
     &   & P$_{sym}$(0.24) & 7.8 \\
     
    & & P$_{snm}$(0.32) & 17.23 \\
    \cline{1-4}
  \end{tabularx}
\end{table*}
  regression. We use these calculated values to compute the NS radius and tidal deformability for any mass M within the above mass range through interpolation for which we do not have the analytical expression.

The values of R\(^2\) are \(\sim\) 0.99,  0.93 and  0.92 for NS radii, tidal deformabilities and maximum mass, respectively. To assess the model prediction we calculate the radii and tidal deformabilities of NS with mass 1.425 M$_\odot$ by interpolating the results from the SRMs. We chose this mass because most of the NSs clustered close to $\sim$ 1.4 M$_\odot$. The Fig~(\ref{fig1}) displays the predictions of the SRMs for R$_{1.425}$, $\Lambda_{1.425}$ and M$_{max}$. The RMSE scores are 0.33 km, 64 and 0.03 M$_\odot$ for R$_{1.425}$, $\Lambda_{1.425}$ and M$_{max}$, respectively. The $\frac{\delta{\mathcal T}}{{\mathcal T}}$ = \(\frac{({\mathcal T}_{\rm TOV}- {\mathcal T}_{\rm SRM})\times100}{{\mathcal T}_{\rm TOV}}\) is the relative percentage error for a given target variable ${\mathcal T}$, plotted as a function of \({\mathcal T}_{\rm TOV}\). Here \({\mathcal T}_{\rm TOV}\) and \({\mathcal T}_{\rm SRM}\) denote the NS property derived using the solutions of TOV equations and using symbolic regression models, respectively. For the case of $\Lambda_{1.425}$, there are a few samples that show a relative error greater than 30$\%$. To analyze this in more detail, in the Fig~\ref{fig2} we plot the variation of L$_0$ with K$_{\rm sym0}$ for these specific cases. The large relative errors ($>$30$\%$) occur for the EoSs with very low K$_{\rm sym0}$($<$-200 MeV) and/or very small L$_0$ ($<$40 MeV). These values of L$_0$ and K$_{\rm sym0}$ are somewhat away from the commonly accepted ranges.

\subsection{Sensitivity of the EoS to Various Data}

We use the SRMs instead of solving TOV equations to expedite our investigation based on Bayesian inferences. We perform Bayesian inference for various scenarios to examine the sensitivity of the EoS to various nuclear and astrophysics inputs. Before embarking on analysis using the realistic set of data, we considered a set of pseudo data as listed in Table~\ref{tab2}. These pseudo data are generated using a suitable set of NMPs referred to as `ref' in Table~\ref{tab3}. The pseudo inputs related to NS are generated through the solutions of TOV equations. We use the empirical data on symmetry energy (e$_{\rm sym}$), symmetry energy pressure (P$_{\rm sym}$) and pressure of symmetric nuclear matter (P$_{\rm snm}$) from various nuclear experiments spanning the density range 0.03 - 0.32 fm$^{-3}$ referred hereafter as ``pseudo experimental data". In addition, we employ the empirical values of energy per nucleon for pure neutron matter e$_{\rm pnm}$ and slope
of the incompressibility coefficients ${\mathcal M}$ at the crossing density ($\rho \sim$ 0.1 fm$^{-3}$ ). The empirical values of e$_{\rm pnm}$ and ${\mathcal M}$ are derived from nuclear masses and iso-scalar giant monopole resonances~\cite{Brown2013b,Khan_Jerome2013}. The significance of e$_{\rm pnm}$ is that it has a contribution from both symmetric nuclear matter and symmetry energy. To analyse the influence of these data on the EoS, we have considered different scenarios as follows: 
 \begin{itemize}
    \item  S$_1$ : Only maximum mass M$_{max}$ of NS.

\item S$_2$ : Radii for 1.4 M$_\odot$ (R$_{1.4}$) and 2.08 M$_\odot$ (R$_{2.08}$) and tidal deformability for 1.4 M$_\odot$ ($\Lambda_{1.4}$) NS along with M$_{max}$.

\item S$_3$ : The empirical values of e$_{\rm pnm}$ and ${\mathcal M}$ at crossing density ($\rho \sim$ 0.1 fm$^{-3}$) along with the astrophysical inputs considered in the scenario S$_2$.

\item S$_4$ : Data from scenario S$_2$ together with ``pseudo experimental data" (see Table~\ref{tab2}).

\item S$_5$ : All the data, including those from scenario S$_4$, as well as data on e$_{\rm pnm}$ and ${\mathcal M}$.
\end{itemize}

{\renewcommand{\arraystretch}{2.0}
\begin{table*}
\caption{Distributions of the NMPs within 68$\%$  confidence interval obtained from Bayesian inference using symbolic regression models. The reference values (`ref') of the NMPs employed to generate the pseudo data of Table~\ref{tab2} used for Bayesian inference are also listed. We have constructed 5 scenarios for different combinations of these data (see text). Among them only the S$_5$ scenario is also performed with the traditional approach of computing EoS and solving TOV equations which is the last row denoted as ${\mathcal T}_{\rm TOV}$. The unit of All the NMPs are in MeV.} \label{tab3}
  \begin{tabular}{lccccccccc}
  \hline
 & Scenario &K$_0$ & Q$_{0}$ & Z$_{0}$ & J$_{0}$ & L$_{0}$ & K$_{\rm sym0}$ & Q$_{\rm sym0}$ & Z$_{\rm sym0}$ \\ [1.3ex]
 \hline
ref & - & 266 & -90 & -996 & 33.17 & 67 & -47 & 458 & -38 \\[1.3ex] 
 \hline
  \multirow{7}{*}
  & S$_1$ &241$_{-58}^{+56}$ & -118$_{-249}^{+228}$ & -50$_{-1782}^{+1929}$ & 32.42$_{-4.94}^{+4.75}$ & 74$_{-26}^{+28}$ & 6$_{-188}^{+177}$  & 936$_{-392}^{+355}$ &  -3$_{-1827}^{+1873}$ \\[1.3ex]

 & S$_2$ &242$_{-57}^{+56}$ & -142$_{-233}^{+251}$ & 19$_{-1968}^{+1919}$ & 32.59$_{-4.59}^{+4.76}$ & 66$_{-14}^{+15}$ & -22$_{-147}^{+158}$  & 877$_{-348}^{+390}$ & 29$_{-1858}^{+1869}$  \\[1.3ex] 

{${\mathcal T}_{\rm SRM}$} & S$_3$ & 259$_{-23}^{+25}$ & -194$_{-181}^{+183}$ & 2$_{-1881}^{+1951}$ & 33.27$_{-1.85}^{+1.78}$ & 67$_{-16}^{+16}$ & -40$_{-152}^{+169}$  & 886$_{-382}^{+397}$ & -1$_{-1989}^{+2017}$  \\[1.3ex]  

 & S$_4$ &256$_{-56}^{+46}$ & -148$_{-266}^{+255}$ & -263$_{-1790}^{+2079}$ & 32.92$_{-1.33}^{+1.36}$ & 65$_{-14}^{+14}$ & -27$_{-79}^{+81}$  & 911$_{-398}^{+380}$ & 197$_{-2056}^{+1844}$  \\[1.3ex] 

& S$_5$ &262$_{-23}^{+21}$ & -167$_{-194}^{+175}$ & -281$_{-1805}^{+2034}$ & 32.97$_{-1.26}^{+1.29}$ & 65$_{-14}^{+14}$ & -26$_{-80}^{+79}$  & 882$_{-387}^{+402}$ & 160$_{-2023}^{+1889}$  \\[1.3ex] 

 \hline
 ${\mathcal T}_{\rm TOV}$& S$_5$ & 262$_{-20}^{+20}$& -203$_{-176}^{+182}$ & -40$_{-1392}^{+1638}$ & 32.74$_{-1.24}^{+1.24}$& 63$_{-13}^{+14}$ &-34$_{-74}^{+76}$ & 845$_{-358}^{+403}$& 339$_{-1691}^{+2120}$ \\[1.3ex] 
 \hline
  \end{tabular}
\end{table*}
In Table~\ref{tab3}, we present the median and 68$\%$ confidence intervals for the NMPs for various scenarios as discussed above. It is apparent that only scenarios involving M$_{max}$, and R$_{1.4}$, R$_{2.08}$ and $\Lambda_{1.4}$ are not enough to reproduce the reference values. We consider the scenario S$_3$ where the data on e$_{\rm pnm}$ and ${\mathcal M}$ at crossing density ($\rho \sim$ 0.1 fm$^{-3}$) are included in addition to those for M$_{\rm max}$, R$_{1.4}$, R$_{2.08}$ and $\Lambda_{1.4}$. The uncertainties for most of the NMPs are reduced. The median values of the NMPs such as K$_0$, J$_0$, L$_0$ and K$_{\rm sym0}$ become closer to their reference values. It is clear that e$_{\rm pnm}$ and ${\mathcal M}$ associated with low density play a significant role in constraining the overall behaviour of the EoS.
\begin{figure}
    \centering
    \includegraphics[width=0.5\textwidth]{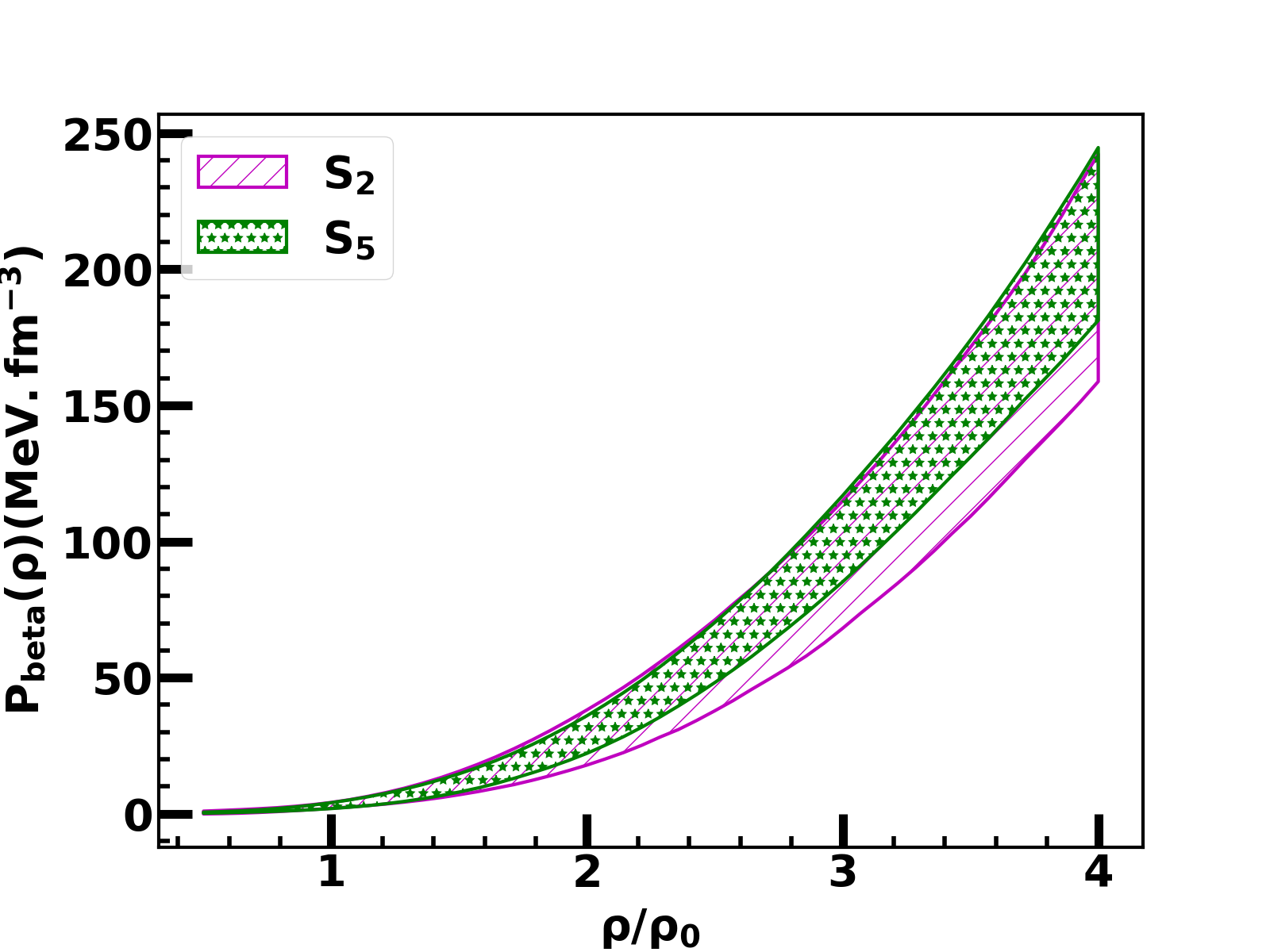}
    \caption{Distribution of $\beta$ equilibrium matter pressure P$_{\rm beta}$ as a function baryon density. S$_2$ represents the scenario where only data from NS observations are considered while S$_5$ represents the scenario in which all the data from low-density nuclear experiments are also considered along with data of S$_2$ scenario.}
    \label{fig3}
\end{figure}
 The scenario S$_4$ which includes all the pseudo data of Table~\ref{tab2} except for e$_{\rm pnm}$ and ${\mathcal M}$, yields smaller uncertainties concerning NMPs associated with symmetry energy compared to the scenario S$_2$, where only the data on NS radii, tidal deformabilities and maximum mass are taken into account. The uncertainty of the K$_0$ is also reduced. When comparing scenarios S$_3$ and S$_4$, it becomes apparent that constraints from e$_{\rm pnm}$ and ${\mathcal M}$ control the symmetric nuclear matter sector of the EoS, while the pseudo experimental data constrains the density-dependent symmetry energy sector of the EoS. Therefore, by combining both sets of data with information on NS radius, tidal deformability and maximum mass in the scenario S$_5$, the distributions of NMPs are found to be consistent with those of S$_3$ and S$_4$. To assess the robustness of our results using SRMs in Bayesian inference, we performed the calculation again only for the scenario S$_5$, by replacing SRMs with the solutions from the TOV equations to compute NS properties. From the last two rows of the table, it is evident that the results are similar in both approaches. Nevertheless, it should be underscored that the SRMs method proves significantly more efficient, performing computations more than 100 times quicker, often within a few minutes. Furthermore, as displayed in Fig~\ref{fig3} the experimental data and empirical values of e$_{\rm pnm}$ and ${\mathcal M}$ have constrained the pressure of $\beta$-equilibrium matter at high density, as evidenced by the results obtained for the scenarios S$_2$ and S$_5$ presented in Table~\ref{tab3}.
\subsection{Constraining the EoS using Realistic Data}
We repeat the analysis using realistic data for the scenario S$_5$, as the reference values for the NMPs are successfully recovered in this case. Bayesian inference is employed with both SRMs and solutions from TOV equations to calculate various properties of neutron stars. The realistic data except for e$_{\rm pnm}$ and ${\mathcal M}$ are exactly the same as in Ref~\cite{Imam2024b}(see Table-I of \cite{Imam2024b}). The e$_{\rm pnm}$ and ${\mathcal M}$ as introduced in the present paper, there values at the crossing density are taken from Refs~\cite{Brown2013b,Khan_Jerome2013}. The SRMs provide the results within 20 minutes being 100 times faster than the approach of computing the EoS and solving TOV equations to calculate the NS properties.

The marginalized posterior distributions (PDs) and the confidence ellipses for the NMPs derived from both SRMs and TOV solutions are displayed as corner plots in Fig~\ref{fig4}. The corner plots exhibit the one-dimensional marginalized PDs for the NMPs on their diagonals. Concurrently, the confidence ellipses are depicted alongside the off-diagonal elements of the corner plots, representing 1$\sigma$, 2$\sigma$, and 3$\sigma$ confidence intervals. The width and inclination of the confidence ellipses for a pair of NMPs determines the nature of the linear correlations among them. It shows L$_0$ is strongly correlated with J$_0$ and K$_{\rm sym0}$ with Pearson's correlation coefficient r $\sim$ 0.9. There is also a stronger correlation between J$_0$ and K$_{\rm sym0}$ (r $\sim$ 0.7). Any other pairs of NMPs do not show significant correlations. It is also clear that the results of Bayesian inference using SRMs and solutions of TOV equations are very similar. 

We also present the posterior distributions of pressure of $\beta$ equilibrium matter (P$_{\rm beta}$($\rho$)), pressure of symmetric nuclear matter (P$_{\rm snm}$($\rho$)) and symmetry energy pressure (P$_{\rm sym}$($\rho$)) in Fig~\ref{fig5}. We observe that the low-density parts of pressure, both for symmetric nuclear matter and symmetry energy, are tightly constrained, and hence the $\beta$-equilibrium matter pressure is also constrained very well at low density ($\rho\leq$ 2$\rho_0$). So the experimental data at low densities effectively limit the symmetry energy within this regime. However, uncertainties grow 
\begin{figure}
\centering
\includegraphics[width=0.44\textwidth]{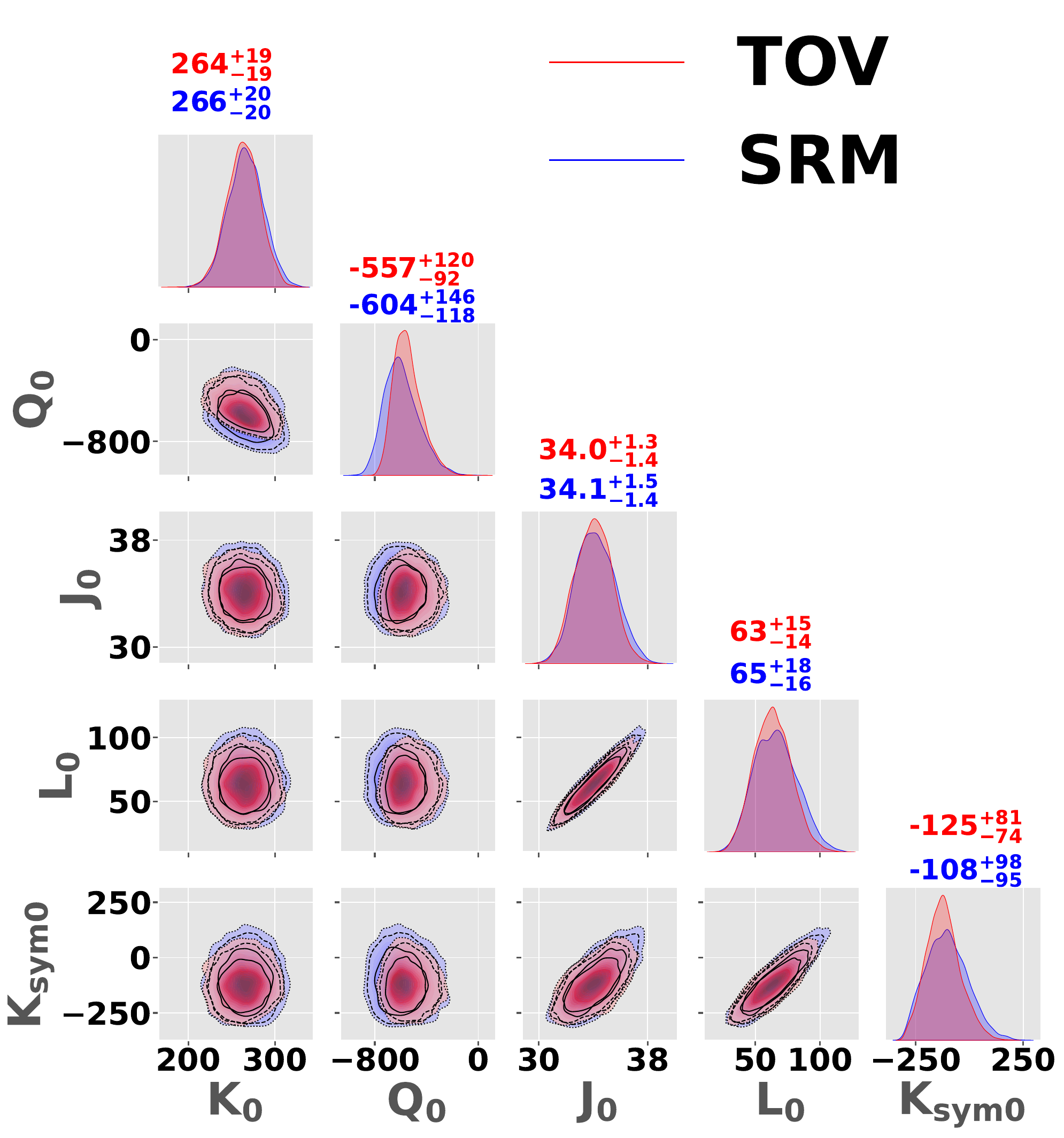}
\caption{\label{fig4}
 The marginalized posterior distributions of the NMPs obtained in the Bayesian inference with the realistic data. We present the results obtained using SRMs (blue) and solutions of TOV equations (red) to compute NS properties in the inference.}
\end{figure}
\begin{figure}[H]
\centering
\includegraphics[width=0.44\textwidth]{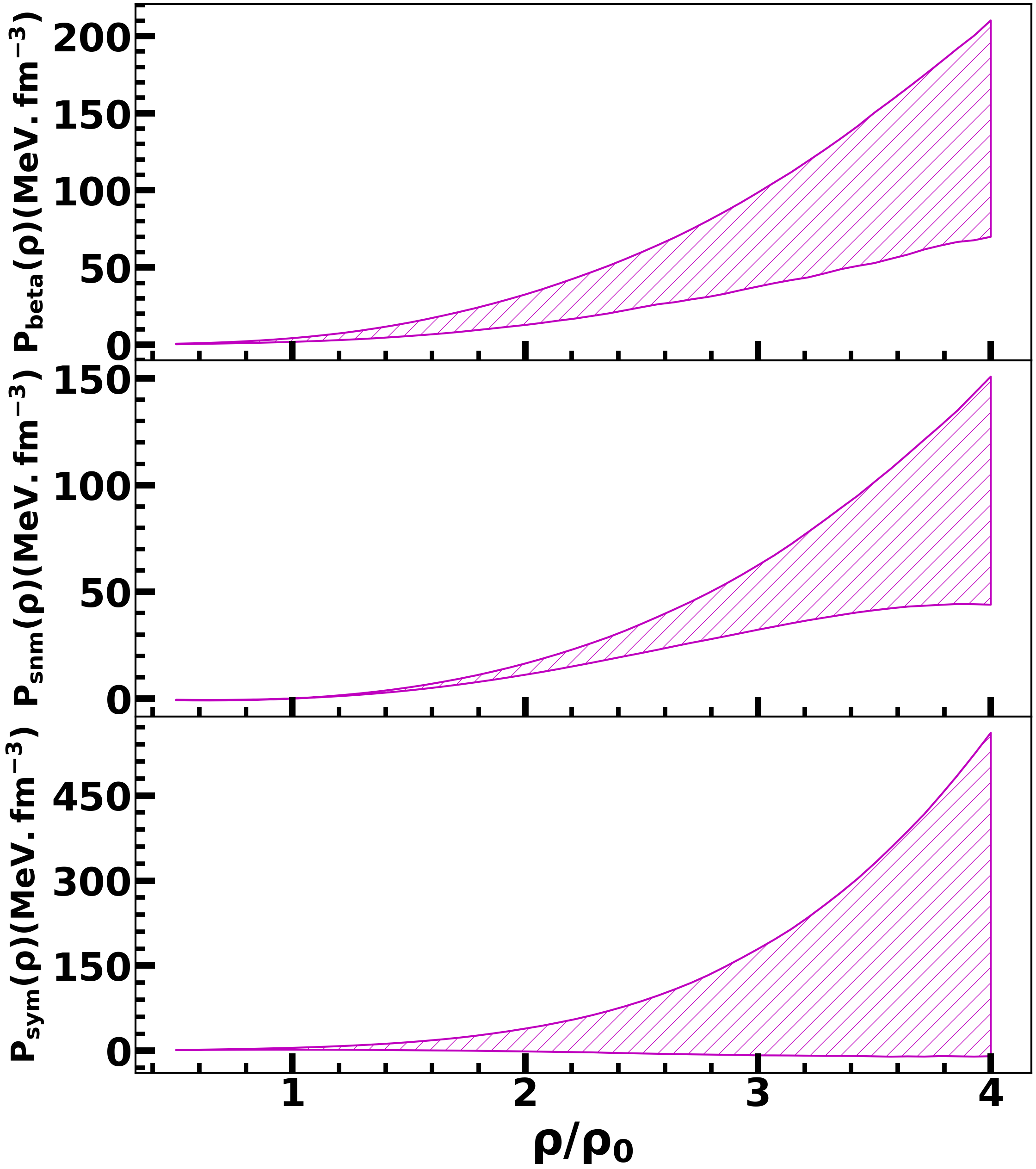}
\caption{\label{fig5}95$\%$ confidence interval for the pressure of $\beta$-equilibrate matter P$_{\rm beta}$($\rho$) (top), pressure of symmetric nuclear matter P$_{\rm snm}$($\rho$)(middle) and symmetry energy pressure P$_{\rm sym}$($\rho$)(bottom) as a function of baryon density. These are calculated using distributions of the NMPs as shown in Fig~\ref{fig4}. The unit of All the NMPs are in MeV.}
\end{figure}
 significantly as we move beyond 2$\rho_0$, especially in the case of the pressure from symmetry energy. This indicates that existing astrophysical data on neutron star properties do not sufficiently constrain symmetry energy at the high-density.
\section{Summary}\label{conc}
We generate a large ensemble of EoSs characterized by the nuclear matter parameters using $\frac{n}{3}$ expansion. The neutron star properties such as radius, tidal defromability and maximum mass corresponding to these EoSs are obtained through the solutions of TOV equations. We have constructed analytical expressions using symbolic regression which map these NS properties to the nuclear matter parameters. For these expressions of radii and tidal deformabilities of NS, we have considered the mass range from 1.1 to 2.15 M$_\odot$, with the interval of 0.05 M$_\odot$. These analytical expressions (SRMs) are employed in Bayesian inference for the faster estimation of NS properties directly in terms of the NMPs. We employed the SRMs in the Bayesian inference to constrain the EoS.

The symbolic regression model approach is computationally faster by more than 100 times compared to those performed with the solutions of TOV equations. This enabled us to perform Bayesian inference for different scenarios to study the sensitivity of various data on the EoS. We have first considered a set of pseudo data for a known value of NMPs so that the accuracy of the posterior distributions of the NMPs can be assessed. We have considered selected data on key EoS quantities and NS properties. The key EoS quantities considered are the pressure for the symmetric nuclear matter, symmetry energy pressure and the symmetry energy, empirically constrained over a range of densities($\sim$0.03 - 0.32 fm$^{-3}$) by the experimental data on bulk properties of finite nuclei and the heavy-ion collision. We have also included the constraints on energy per pure neutron matter and slope of incompressibility coefficient at the crossing density ($\sim$ 0.1 fm$^{-3}$). The astrophysics data considered are the radii for NS with mass $\sim$ 1.4 M$_\odot$ and 2.08 M$_\odot$ and the tidal deformability for 1.4 M$_\odot$ NS and the NS maximum mass.
Our analysis reveals that the key EoS quantities associated with low densities in addition to the NS observations play a significant role in confining the overall behaviour of the EoS (see Fig~\ref{fig3}).\\
\vspace{-0.10cm}
Finally, we have performed Bayesian inference using the aforementioned data obtained from nuclear experiments and astrophysical observations. The symbolic regression models are employed in the Bayesian inference to compute the NS properties. To verify the reliability of our findings with symbolic regression models, we repeated the analysis using the solutions of TOV equations to compute the NS properties. Remarkably, both approaches yielded comparable results, while the SRMs deliver outcomes in just 20 minutes, making them over 100 times faster than the approach of using the solutions of TOV equations within the same computational setup.
\section{Acknowledgements} 
BKA acknowledges partial support from the SERB, Department of Science and Technology, Government of India with  CRG/2021/000101. The authors sincerely acknowledge the usage of the analysis software {\tt BILBY} \cite{Ashton2019,Bilby_ref}.
T.M has received support through national funds from FCT (Fundação para a Ciência e a Tecnologia, I.P, Portugal) for projects UIDB/04564/2020 and UIDP/04564/2020, identified by DOIs 10.54499/UIDB/04564/2020 and 10.54499/UIDP/04564/2020, respectively, and for project 2022.06460.PTDC with the DOI identifier 10.54499/2022.06460.PTDC. T.M would additionally like to express gratitude for the RNCA (Rede Nacional de Computação Avançada) advanced computing projects 2024.14108.CPCA.A3, which received funding from the FCT.

\bibliographystyle{apsrev4-2}
%
\widetext
\newpage
\section*{Appendix}
\subsection*{Equations for Tidal Deformability of NS}
To find the tidal deformability at any mass M between 1.1 M$_\odot$ to 2.15 M$_\odot$ first we compute the tidal deformabilities using the expressions $\Lambda_{\rm M}$ given below for all the masses, then we do interpolate between this mass range. Here each of the NMPs $\theta$ are expressed as a reduced variable ${\hat \theta} = \frac{\theta-\bar{\theta}}{\sigma_\theta}$ with ${\bar \theta}$ and $\sigma_\theta$ are the average and standard deviation of each NMPs as described earlier.
\bea
\Lambda_{1.1} &=&  {\hat L}_0(-118.62{\hat J}_0 - 368.19(\frac{{\hat J}_0}{{\hat L}_0} + {\hat K}_{\rm sym0} - {\hat L}_0) + 1194.33) + 2696.03\\
 \Lambda_{1.15} &=& {\hat L}_0((201.40 - {\hat L}_0)(-{\hat J}_0 - \frac{{\hat J}_0}{{\hat L}_0} - {\hat K}_{\rm sym0} + {\hat L}_0) + 845.04) + 2197.35 \\
\Lambda_{1.2} &=&  -96.86({\hat J}_0 - {\hat K}_{\rm sym0} - {\hat Q}_{\rm sym0} + {\hat Z}_0) + {\hat L}_0({\hat L}_0 + 582.78)  + 1835.62\\
 \Lambda_{1.25} &=&  -100.89({\hat J}_0 - {\hat K}_{\rm sym0}  -{\hat Q}_{\rm sym0}) + {\hat L}_0(410.36 - {\hat K}_{\rm sym0}) - 54.39{\hat Z}_0 + 1445.51\\
 \Lambda_{1.3} &=&   -51.27({\hat J}_0 - {\hat K}_0 - {\hat Q}_0 -{\hat Q}_{\rm sym0}) + 102.54{\hat K}_{\rm sym0} + 296.27{\hat L}_0  + 1160.47\\
\Lambda_{1.35}& = &  -53({\hat J}_0 - {\hat K}_0 - {\hat Q}_0 -{\hat Q}_{\rm sym0}) + 108.45({\hat K}_{\rm sym0} + 2{\hat L}_0)  + 919.09\\ 
\Lambda_{1.4} &=&  -54.02({\hat J}_0 - {\hat K}_0 - {\hat Q}_0 -{\hat Q}_{\rm sym0}) + 76.09{\hat K}_{\rm sym0} + 178.12{\hat L}_0  + 737.93\\
\Lambda_{1.45} &=&  -40.17({\hat J}_0 - {\hat Q}_0 - {\hat Q}_{\rm sym0}) + 48.08({\hat K}_0 + {\hat K}_{\rm sym0}) + 136.60{\hat L}_0  + 595.79\\
\Lambda_{1.5} &=&  -37.67({\hat J}_0 - {\hat K}_0) + 82.07({\hat K}_{\rm sym0} + {\hat L}_0) + 26.79({\hat Q}_0 + {\hat Q}_{\rm sym0}) + 484.31\\
 \Lambda_{1.55} &=& -25.89({\hat J}_0 - {\hat Q}_{\rm sym0}+{\hat Z}_0) + 27.54{\hat K}_0 + 63.88({\hat K}_{\rm sym0} + {\hat L}_0)  + 391.95\\ 
 \Lambda_{1.6} &=&  -21.54({\hat J}_0 -{\hat Q}_0 - {\hat Q}_{\rm sym0}) + 43.08{\hat K}_0 + 53.01({\hat K}_{\rm sym0} + {\hat L}_0)  + 319.95\\
\Lambda_{1.65} &= & (-{\hat K}_0 - 20.51)({\hat J}_0 - {\hat K}_0 - 2.14{\hat K}_{\rm sym0} - 2.14{\hat L}_0 - {\hat Q}_0 - {\hat Q}_{\rm sym0} - 12.75)\\
\Lambda_{1.7} &=&  -12.47({\hat J}_0 - {\hat Q}_0 - {\hat Q}_{\rm sym0}) + 24.93{\hat K}_0 + 34.31({\hat K}_{\rm sym0} + {\hat L}_0)  + 213.74\\
\Lambda_{1.75} &=&  -11.12{\hat J}_0 + 24.89({\hat K}_0 + {\hat K}_{\rm sym0} + {\hat L}_0) + 14.95({\hat Q}_0 + {\hat Q}_{\rm sym0}) + 174.25\\
\Lambda_{1.8} &= & -10.24({\hat J}_0 - {\hat Q}_0 - {\hat Q}_{\rm sym0})  + 18.04({\hat K}_0 + {\hat L}_0) + 25.83{\hat K}_{\rm sym0}  + 142.42\\
\Lambda_{1.85} &=& -8.05{\hat J}_0 + 19.96({\hat K}_0 + {\hat K}_{\rm sym0}) + 12({\hat L}_0 + {\hat Q}_0 + {\hat Q}_{\rm sym0}) + 116.10\\ 
\Lambda_{1.9} &=& -8.64({\hat J}_0 - {\hat L}_0 - {\hat Q}_0 - {\hat Q}_{\rm sym0}) + 15.10{\hat K}_0 + 17.28{\hat K}_{\rm sym0}  + 94.47\\
\Lambda_{1.95} &=&  -6.42({\hat J}_0 - {\hat L}_0 - {\hat Q}_{\rm sym0}) + 14.69({\hat K}_0 + {\hat K}_{\rm sym0})  + 9.20{\hat Q}_0  + 76.57\\
\Lambda_{2.0}& = & -4.78({\hat J}_0 - {\hat L}_0 - {\hat Q}_{\rm sym0} - {\hat Z}_0) + 12.84({\hat K}_0 + {\hat K}_{\rm sym0}  + {\hat Q}_0) + 61.66 \\
\Lambda_{2.05} &=&  -4.32({\hat J}_0 - {\hat Q}_{\rm sym0}) + 11.34{\hat K}_0 + 8.38({\hat K}_{\rm sym0} - {\hat Q}_0) + {\hat L}_0  + 49.33\\
\Lambda_{2.1} &=&  -3.57({\hat J}_0 - {\hat Q}_{\rm sym0}) + 11.23{\hat K}_0 + 7.66({\hat K}_{\rm sym0} + {\hat Q}_0)  + {\hat Z}_{\rm sym0} + 38.89\\
\Lambda_{2.15} &=&  {\hat K}_0 + (-{\hat K}_0 - {\hat Q}_0 - 3.88)(0.67{\hat J}_0 - 1.32{\hat K}_{\rm sym0} - {\hat Q}_{\rm sym0} - 7.76) 
\eea
\subsection*{Equations for Radius of NS}
To calculate the radius of a neutron star at any mass M we have to follow the same procedure as discussed about tidal deformability in the previous section.
\bea
R_{1.1}  &=& -0.26({\hat J}_0 - {\hat Q}_{\rm sym0}-{\hat Z}_0) - 1.26{\hat K}_{\rm sym0} + 2.42{\hat L}_0 + {\hat Q}_0 + 14.76\\
R_{1.15} &=&  -0.38({\hat J}_0 + {\hat K}_0 - {\hat Q}_{\rm sym0} +{\hat Z}_0) - 1.02{\hat K}_{\rm sym0} + 2.30{\hat L}_0   + 14.66\\
R_{1.2}  &=&  -0.36({\hat J}_0 + {\hat K}_{\rm sym0}{\hat L}_0 -{\hat Q}_0 - {\hat Q}_{\rm sym0}) - {\hat K}_{\rm sym0} + 2{\hat L}_0  + 14.55\\
R_{1.25} &=&  -0.33({\hat J}_0 - {\hat Q}_0 - {\hat Q}_{\rm sym0})- 0.18{\hat K}_0 - {\hat K}_{\rm sym0} + 2{\hat L}_0  + 14.43\\
R_{1.3}  &=&  -0.21({\hat J}_0 + {\hat K}_0 - {Q}_0 - {\hat Q}_{\rm sym0}) - 0.79K_{\rm sym0} + 2{\hat L}_0 + 14.36\\
R_{1.35} &=& -0.30(\hat{J}_0 - \hat{Q}_0 - \hat{Q}_{\rm sym0}) - 0.70\hat{K}_{\rm sym0} + 1.70\hat{L}_0  + 14.28 \\
R_{1.4}  &=&-0.31({\hat J}_0 - {\hat Q}_0 - {\hat Q}_{\rm sym0}) -0.62{\hat K}_{\rm sym0} + 1.75{\hat L}_0  + 14.22\\  
R_{1.45}&=& {\hat L}_0 +14.15 -(0.51({\hat J}_0  - {\hat L}_0 - {\hat Q}_0 - {\hat Q}_{\rm sym0}) + {\hat K}_{\rm sym0})/(2.37 - {\hat L}_0) \\ 
R_{1.5} &=& -0.25(\hat{J}_0 - \hat{Q}_{\rm sym0}) - 0.43(\hat{K}_{\rm sym0} - \hat{Q}_0) + 1.43\hat{L}_0  + 14.08 \\ 
R_{1.55} &=& -0.22(\hat{J}_0 - \hat{Q}_0 - \hat{Q}_{\rm sym0}) - 0.45\hat{K}_{\rm sym0} + 1.38\hat{L}_0  + 14.03 \\  
R_{1.6} &=& -0.25(\hat{J}_0 + \hat{K}_{\rm sym0}\hat{L}_0 + \hat{K}_{\rm sym0} - \hat{Q}_0 - \hat{Q}_{\rm sym0}) + 1.25\hat{L}_0  + 13.97 \\
R_{1.65} &=& \hat{L}_0 + 13.92 - (0.63\hat{J}_0 + \hat{K}_{\rm sym0} - \hat{L}_0 - \hat{Q}_0 - \hat{Q}_{\rm sym0})/(3.95 - \hat{L}_0) \\
R_{1.7} &=& -0.23(\hat{J}_0 - \hat{Q}_{\rm sym0}) - 0.27(\hat{K}_{\rm sym0} - \hat{Q}_0) + 1.17\hat{L}_0  + 13.85 \\
R_{1.75} &=& -0.23(\hat{J}_0 + \hat{K}_{\rm sym0} - \hat{Q}_0 -\hat{Q}_{\rm sym0}) - 0.17\hat{K}_{\rm sym0}\hat{L}_0  + \hat{L}_0  + 13.81 \\
R_{1.8} &=& -0.21(\hat{J}_0 + \hat{K}_{\rm sym0} - \hat{Q}_{\rm sym0} + \hat{Z}_0) - 0.16\hat{K}_{\rm sym0}\hat{L}_0 + \hat{L}_0  + 13.74 \\ 
R_{1.85} &=& -0.20(\hat{J}_0 + \hat{K}_{\rm sym0} - \hat{Q}_{\rm sym0}) + 0.13\hat{K}_0 + \hat{L}_0 + 0.32\hat{Q}_0 + 13.67 \\
R_{1.9} &=& -0.16(\hat{J}_0 - \hat{K}_0 + \hat{K}_{\rm sym0}) + \hat{L}_0 + 0.32\hat{Q}_0 + 0.24\hat{Q}_{\rm sym0} + 13.58 \\
R_{1.95} &=& -0.18(\hat{J}_0 - \hat{K}_0 + \hat{K}_{\rm sym0} -2\hat{Q}_0 - \hat{Q}_{\rm sym0}) + 0.82\hat{L}_0  + 13.52 \\
R_{2.0} &=& -0.18\hat{J}_0 + 0.24\hat{K}_0 + 0.76\hat{L}_0 + 0.34\hat{Q}_0 + 0.24\hat{Q}_{\rm sym0} + 13.43 \\
R_{2.05} &=& -0.15\hat{J}_0 + 0.24\hat{K}_0 + 0.69\hat{L}_0 + 0.31\hat{Q}_0 + 0.24\hat{Q}_{\rm sym0} + 13.37 \\
R_{2.1} &=& -0.16\hat{J}_0 + 0.31\hat{K}_0 + 0.61\hat{L}_0 + 0.36\hat{Q}_0 + 0.37\hat{Q}_{\rm sym0} + 13.21 \\
R_{2.15} &=& -0.19\hat{J}_0 + 0.28\hat{K}_0 + 0.57\hat{L}_0 + 0.38\hat{Q}_0 + 0.28\hat{Q}_{\rm sym0} + 13.06
\eea
\subsection*{Equation for Maximum Mass of Neutron Star}
The equation for the maximum mass of a neutron star is given as,
\bea
M_{\rm max} &=& 0.14{\hat K}_0 - 0.02{\hat L}_0 + 0.27{\hat Q}_0 + 0.16{\hat Z}_0 + 2.33\\ \nonu
\eea

\end{document}